# What Matters Most? A Quantitative Meta-Analysis of AI-Based Predictors for Startup Success

Seyed Mohammad Ali Jafari[1]

Ali Mobini Dehkordi[2]

Ehsan Chitsaz[3]

Yadollah Yaghoobzadeh[4]

## Abstract

**Background:** Predicting startup success with machine learning is a rapidly growing field, yet findings on key predictors are often fragmented and context-specific. This makes it difficult to discern robust patterns and highlights a need for a systematic synthesis of the evidence.

**Methods:** This study conducts a quantitative meta-analysis to synthesize the literature on predictor importance in AI-based startup evaluation. We performed a systematic review to identify a final sample of 13 empirical studies that report rankable feature importance. From these papers, we extracted and categorized 58 unique predictors, synthesizing their importance using a Weighted Importance Score (WIS) that balances a feature's average rank with its frequency of appearance. We also conducted a moderator analysis to investigate how predictor importance changes with context (e.g., success definition, startup stage).

**Results:** Our aggregate analysis reveals that the most consistently powerful predictors are a quartet of foundational attributes: Firm Characteristics (e.g., age, location), Investor Structure (e.g., investor quality), Digital and Social Traction (e.g., online momentum), and Funding History. The moderator analysis further reveals that this hierarchy is highly context-dependent. For instance, predicting near-term funding milestones elevates the importance of the deal's immediate context, while predicting long-term exits prioritizes fundamental firm and investor characteristics.

**Conclusion:** The factors that best predict startup success are not universal but are contingent on the startup's goals, stage, and the data used for evaluation. Our findings point to a potential "convenience bias" in the literature, where predictor importance may be tied to data accessibility. We conclude by underscoring the need for standardized reporting practices to enable more robust, cumulative knowledge building in the field.

**Keywords:** Startup Success, Feature Importance, Meta-Analysis, Machine Learning, Artificial Intelligence, Venture Capital, Entrepreneurship.

[1] PhD candidate, Technological Entrepreneurship Department, Faculty of Entrepreneurship, University of Tehran, Iran, sma_jafari@ut.ac.ir
[2] Full Professor, Faculty of Entrepreneurship, University of Tehran, Iran, mobini@ut.ac.ir
[3] Associate Professor, Faculty of Entrepreneurship, University of Tehran, Iran, chitsaz@ut.ac.ir
[4] Assistant Professor, School of Electrical and Computer Engineering, University of Tehran, Tehran, Iran, y.yaghoobzadeh@ut.ac.ir

# 1. Introduction

Startups are a critical engine of the modern economy, driving innovation and reshaping entire industries. However, the process of Early Stage venture investment remains fraught with uncertainty. For founders seeking capital and investors deploying it, the notoriously high failure rate of new ventures—with some estimates as high as 78%—presents a significant and persistent challenge (Antretter et al., 2019). In response, a growing number of researchers have turned to artificial intelligence and machine learning, seeking to apply data-driven rigor to this high-stakes decision-making process. Leveraging large-scale venture databases like Crunchbase, studies have increasingly applied computational models to identify objective, quantifiable signals that correlate with a startup's future success (e.g., Kim et al., 2023; Żbikowski & Antosiuk, 2021).

This burgeoning field of research has produced a valuable but fragmented body of knowledge. Numerous empirical studies have developed predictive models, yet their findings on which features are most important are often context-specific and appear contradictory. For instance, one study may find that founder experience is paramount (Ferry et al., 2018), while another points to the power of online legitimacy measured via social media activity (Antretter et al., 2019), and a third highlights the primary importance of geographic context (Żbikowski & Antosiuk, 2021). This proliferation of individual, often conflicting, findings has created a "forest for the trees" problem for both academics and practitioners. It is difficult to discern which factors are most consistently predictive across the diverse landscape of startups and investment contexts, creating a clear need for a systematic synthesis to identify the most robust and generalizable patterns.

To address this gap, this study conducts a quantitative meta-analysis of the feature importance literature in AI-based startup prediction. We began by systematically reviewing the field to identify all relevant empirical studies, and from this corpus, we isolated a final sample of 13 studies that provide rankable, quantitative evidence on predictor importance. By aggregating and analyzing the findings from these papers, this research aims to achieve three primary objectives. First, we establish an evidence-based hierarchy of predictor families to answer the question: Overall, which categories of predictors are most influential? Second, we investigate how predictor importance changes with context by conducting a moderator analysis to answer: How does what matters most depend on the startup's goals, stage, and the data used to evaluate it? Third, through this process, we provide a meta-scientific critique of the field, identifying systemic issues like convenience bias and look-ahead bias, topics of growing concern in the recent literature (e.g., Park et al., 2024), and propose a path toward more rigorous and standardized research practices.

The remainder of this paper is structured as follows. Section 2 details our multi-stage methodology, including the systematic review and the criteria for inclusion in the meta-analysis. Section 3 presents the quantitative results of our meta-analysis, including both the overall predictor rankings and the moderator analysis. Section 4 discusses the theoretical and practical implications of these findings. Finally, Section 5 concludes with a summary of our contributions, an acknowledgement of the study's limitations, and an agenda for future research.

# 2. Methodology

To build a comprehensive evidence base for this meta-analysis, we adopted a multi-stage methodology designed for maximum rigor and transparency. The process followed the PRISMA 2020 guidelines for systematic reviews (Page et al., 2021) and is visually summarized in Figure 1.

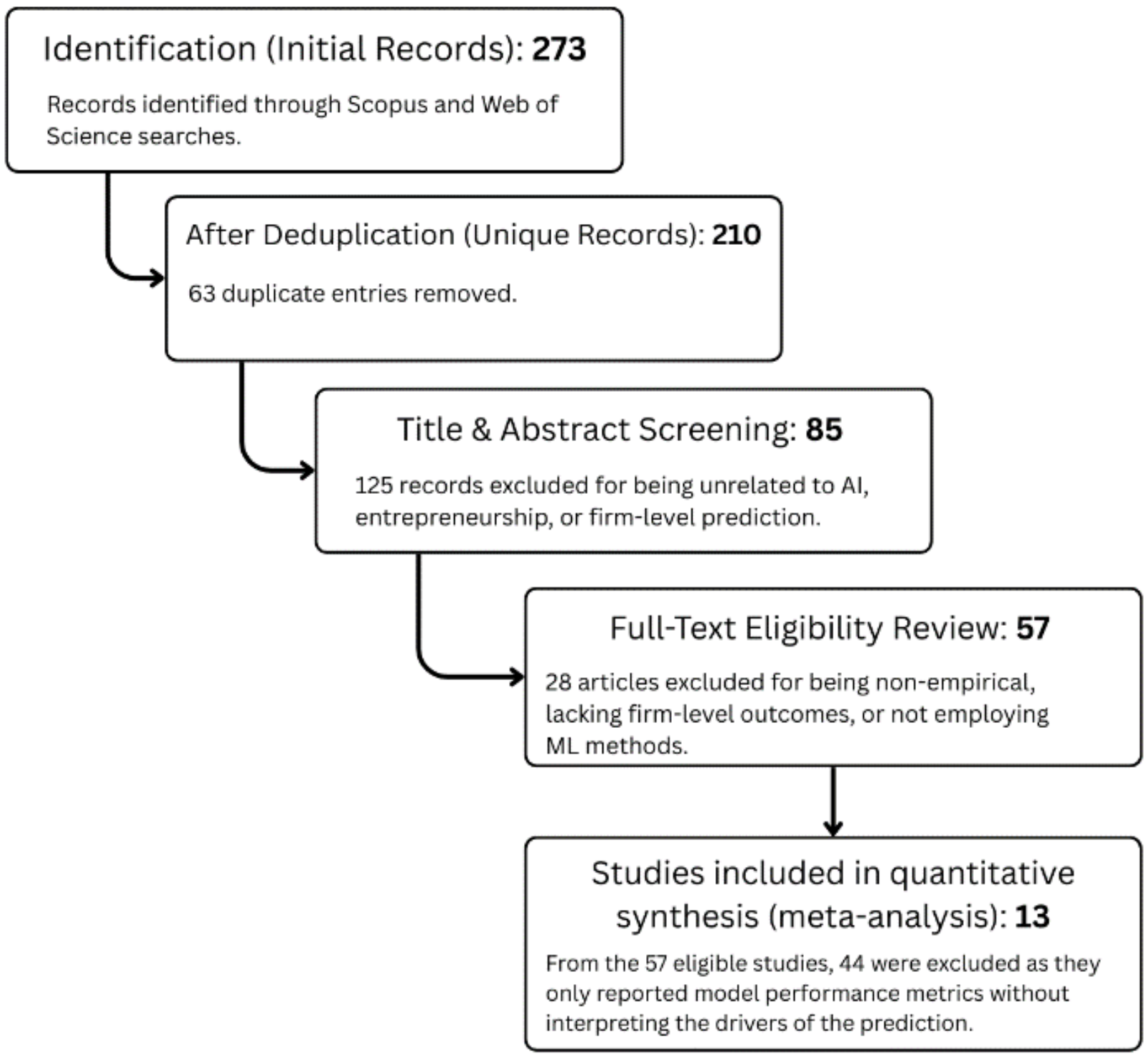

**Figure 1**. PRISMA 2020 flow diagram illustrating the systematic study selection process. The diagram shows the flow of information through the different phases of the review, from initial record identification to the final sample of 13 studies included in the quantitative meta-analysis. Adapted from Page et al. (2021).

### 2.1. Search Strategy and Databases

The search process was designed to identify all relevant empirical studies at the intersection of startup evaluation and artificial intelligence. We conducted our search in July 2025 across two leading academic databases, Scopus and Web of Science, chosen for their extensive coverage of peer-reviewed literature in the relevant subject areas.

The final search string was constructed from four core conceptual blocks to ensure high precision and recall:

```
( "forecasting" OR "prediction" OR "classification" OR "prognosis" OR
"evaluation" ) AND ( "startup" OR "start-up" OR "new venture" OR
"entrepreneurship" ) AND ( "success" OR "survival" OR "failure" OR
"performance" ) AND ("Artificial Intelligence" OR "AI" OR "machine
learning" OR "ml" OR "deep learning")
```

This query was applied to the title, abstract, and keywords of articles in the databases, with no starting date restriction to capture the full history of the field.

## 2.2. Study Selection and Screening

### 2.2.1. Inclusion and Exclusion Criteria

To define the scope of the review and ensure the relevance of the final corpus, we established a strict set of inclusion and exclusion criteria. The review was limited to peer-reviewed articles published in English. A study was included if it met all of the following conditions:

- It presented an empirical study involving data analysis.
- It developed or tested a computational model using AI, machine learning, or related data science techniques.
- Its prediction target was related to firm-level success (e.g., survival, exit, funding, performance).

A study was excluded if it was a non-empirical work (e.g., literature review, editorial, conceptual-only paper), if its primary focus was not on firm-level outcomes, or if it did not use AI/ML techniques for prediction. A key decision in our protocol was to include both peer-reviewed journal articles and premier conference proceedings to ensure our review captures the full spectrum of current, high-quality research in the fast-paced AI field.

### 2.2.2. Screening Process

The screening process followed a four-step funnel, as illustrated in Figure 1.

1. **Identification:** The initial search across Scopus and Web of Science yielded 273 records.
2. **Deduplication:** After combining the results, 63 duplicate entries were removed, leaving 210 unique articles for screening.
3. **Title and Abstract Screening:** The titles and abstracts of these articles were screened against the inclusion criteria. This led to the exclusion of 125 articles that were clearly not focused on AI-based startup success prediction.
4. **Full-Text Eligibility Review:** The remaining 85 articles underwent a full-text review. At this stage, a further 28 articles were excluded for being non-empirical, having an irrelevant outcome variable, or not employing an AI/ML methodology. This process resulted in a corpus of 57 relevant empirical studies.

**2.2.3. Final Selection for Meta-Analysis**
The primary goal of this study is to quantitatively synthesize the *relative importance* of different predictors. Therefore, we applied one final selection criterion to the 57 studies: a study was included in our meta-analytic sample only if it reported a quantitative, rankable measure of feature importance (e.g., a SHAP plot, table of Gini scores, etc.). From the 57 eligible studies, 44 were excluded as they only reported model performance metrics without interpreting the drivers of the prediction.

This final step resulted in our targeted sample of N=13 studies, which forms the complete evidence base for the quantitative analysis presented in this paper.

# 3. Results

This section details the findings of our quantitative meta-analysis. We first provide a descriptive overview of the studies that constitute our meta-analytic sample, establishing the evidence base for our subsequent analysis. We then present the core results concerning the overall importance of different predictor categories, before proceeding to a moderator analysis that explores how this hierarchy of importance varies across different study contexts.

## 3.1. Descriptive Statistics of the Meta-Dataset

To construct a robust evidence base for our meta-analysis, we systematically reviewed the literature and identified 13 empirical studies published between 2018 and 2024 that reported quantitatively rankable feature importance for predicting startup success. These specific studies, which form the final sample for our analysis, provide the raw data on predictor influence from which our aggregate findings are synthesized.

To characterize the landscape of this research, Table 1 provides a breakdown of these 13 studies across five key methodological dimensions: the geographic focus of their dataset, their definition of success, the primary data source employed, the startup stage analyzed, and the scale of the dataset.

**Table 1:** Descriptive Characteristics of Studies Included in the Meta-Analysis (N=13)

| DIMENSION | CATEGORY | N | PCT. | INCLUDED STUDIES (BY FIRST AUTHOR) |
|---|---|---|---|---|
| GEOGRAPHIC FOCUS | Global | 4 | 30.8% | Gangwani (2023), Kim (2023), Razaghzadeh Bidgoli (2024), Żbikowski (2021) |
| | USA / North America | 1 | 7.7% | Ferry (2018) |
| | EU | 1 | 7.7% | Antretter (2019) |
| | Asia | 5 | 38.5% | Cholil (2024), Deodhar (2024), Gautam (2024), Hu (2022), Gujarathi (2024) |
| | Other (e.g., Iran) | 2 | 15.3% | Moterased (2021), Eljil (2024) |
| SUCCESS OUTCOME | Exit Event | 5 | 38.5% | Cholil (2024), Gangwani (2023), Kim (2023), Razaghzadeh Bidgoli (2024), Żbikowski (2021) |
| | Funding Milestone | 3 | 23.1% | Deodhar (2024), Dziubanovska (2024), Eljil (2024) |
| | Survival / Lifespan | 1 | 7.7% | Antretter (2019) |
| | Valuation / Return | 2 | 15.4% | Ferry (2018), Hu (2022) |
| | Other / Ambiguous | 2 | 15.4% | Gautam (2024), Gujarathi (2024) |
| PRIMARY DATA SOURCE | Venture Database | 6 | 46.2% | Gangwani (2023), Gautam (2024), Hu (2022), Kim (2023), Razaghzadeh Bidgoli (2024), Żbikowski (2021) |
| | Bespoke (Web Scrape, Survey, TV Show, etc.) | 7 | 53.8% | Antretter (2019), Cholil (2024), Deodhar (2024), Dziubanovska (2024), Eljil (2024), Ferry (2018), Moterased (2021) |
| STARTUP STAGE | Explicitly Early Stage | 7 | 53.8% | Antretter (2019), Deodhar (2024), Dziubanovska (2024), Eljil (2024), Ferry (2018), Moterased (2021), Żbikowski (2021) |
| | General / Mixed-Stage | 6 | 46.2% | Cholil (2024), Gangwani (2023), Gautam (2024), Hu (2022), Kim (2023), Razaghzadeh Bidgoli (2024) |
| DATASET SCALE | Large-N (>10,000) | 4 | 30.8% | Gangwani (2023), Gautam (2024), Kim (2023), Żbikowski (2021) |
| | Small-N (<1,000) | 9 | 69.2% | Antretter (2019), Cholil (2024), Deodhar (2024), Dziubanovska (2024), Eljil (2024), Ferry (2018), Hu (2022), Moterased (2021), Razaghzadeh Bidgoli (2024) |

As shown in Table 1, the research providing feature importance data is geographically diverse. The largest cluster of studies (38.5%) focuses on Asian markets, including China and India. This is followed by a significant number of studies (30.8%) that utilize Global datasets. This broad geographic distribution ensures our analysis captures a worldwide perspective on startup success prediction.

A critical dimension of our meta-dataset is the heterogeneity in how "startup success" is operationalized. The most common definition, used in 38.5% of the studies, is a successful Exit Event, such as an IPO or acquisition. This is followed by securing a future Funding Milestone (23.1%), reflecting the different lenses through which success can be viewed. This definitional variance is a key contextual factor we explore in our moderator analysis.

Methodologically, the field is evenly split regarding data sources. A slight majority of studies (53.8%) rely on Bespoke Datasets curated through methods like web scraping or surveys, while a substantial portion (46.2%) is anchored in large-scale Venture Databases like Crunchbase.

Crucially for our moderator analysis, the sample is also well-balanced across Startup Stage and Dataset Scale. A majority of studies (53.8%) explicitly focus on the Early Stage, while the remainder analyze a broader, mixed-stage population. Furthermore, a clear distinction exists between Large-N studies (30.8%) that analyze over 10,000 companies and Small-N studies (69.2%) that use smaller, more curated datasets. These two dimensions will serve as powerful moderators for dissecting our primary findings.

Having characterized the methodological landscape of our source studies, we now proceed to the central analysis of this paper: the aggregation of feature importance across these diverse contexts.

## 3.2. Core Finding: Overall Predictor Importance

To answer our first research question—"Overall, which categories of predictors are most influential?"—we aggregated the feature importance data from all 13 studies in our meta-analytic sample. Our methodology was designed to synthesize findings from diverse studies into a single, comparable framework.

Our primary method relies on the normalized rank of each feature within its source paper. We first categorized each of the 58 individual predictors identified across the 13 studies into one of eight standardized "Predictor Families." To provide complete transparency on this critical categorization step, Table 2 presents a detailed thematic typology of these families, defining each core concept and listing representative examples of the specific features assigned to it from the source literature. The complete ledger linking every feature to its family is detailed in Appendix B.

**Table 2:** Thematic Typology of Predictor Families

| PREDICTOR FAMILY | CORE CONCEPT & REPRESENTATIVE FEATURES FROM THE LITERATURE |
|---|---|
| FIRM CHARACTERISTICS | Fundamental, often static, attributes of the venture itself. *(e.g., Company age, Country/Region of HQ, Number of employees, Industry category, Crunchbase rank)* |
| FUNDING HISTORY | The magnitude, timing, and velocity of capital acquisition. *(e.g., Total funding amount, Last raised amount, Time since last funding, Number of funding rounds, Funding goal, Ask equity)* |
| TEAM/FOUNDER | The human capital, experience, and composition of the founding team. *(e.g., Founder years of experience, Founder prior exits, University rank, Team size, Number of advisors, Educational background)* |
| INVESTOR STRUCTURE | The quality, composition, and strategic behavior of a startup's financial backers. *(e.g., Investor's historical success rate, Number of investors, Investment size, Number of limited partners, Presence of top-tier VCs)* |
| DIGITAL AND SOCIAL TRACTION | Proxies for market presence, brand momentum, and user engagement derived from online sources. *(e.g., LinkedIn/Twitter followers, Website mentions & activity, User engagement score, Length of tweets)* |
| MARKET/SECTOR | The industrial and competitive context in which the startup operates. *(e.g., Industry category, Market size, Competition level, Industry convergence)* |
| PSYCHOMETRIC/BEHAVIORAL | Direct measures of founder personality, attitudes, and cognitive traits. *(e.g., Entrepreneurial intention, Fear of failure, Opportunity perception, Self-efficacy)* |
| PRODUCT/TECH | The maturity and defensibility of the startup's core technology or product. *(e.g., Has MVP, Patent counts, Quality of the idea, Bag-of-words from description)* |

With this clear classification scheme, we then measured the overall importance of each family using two complementary metrics: (1) Frequency of Appearance, which counts how often a family's features appear in the top-ranked positions (Top 3 and Top 5), and (2) Average Rank, the arithmetic mean of all ranks for features within that family.

The results of this synthesis are presented in Table 3, while Table 3 provides the full quantitative ranking ordered from most to least important based on the definitive WIS.

To answer our first research question—"Overall, which categories of predictors are most influential?"—we aggregated the feature importance data from all 13 studies in our meta-analytic sample. To create a robust metric that accounts for both the average importance (rank) and the volume of evidence (feature count), we developed a Weighted Importance Score (WIS), calculated as `(1 / Average Rank) * ln(1 + Feature Count)`. A higher WIS indicates a more significant predictor family. Table 2 defines the predictor families, and Appendix B contains the full data ledger.

The overall results of this synthesis are presented in Table 3, which ranks the predictor families from most to least important based on the definitive WIS.

**Table 3:** Aggregated Importance of Predictor Families, Ordered by Weighted Importance Score (WIS) (N=13)

| RANK | PREDICTOR FAMILY | WIS | AVERAGE RANK | COUNT IN TOP 5 |
|---|---|---|---|---|
| **1** | **Firm Characteristics** | 0.93 | 2.75 | 12 |
| **2** | **Investor Structure** | 0.92 | 2.60 | 10 |
| **3** | **Digital and Social Traction** | 0.90 | 2.56 | 9 |
| **4** | **Funding History** | 0.86 | 2.80 | 11 |
| **5** | **Team/Founder** | 0.75 | 3.20 | 10 |
| **6** | **Market/Sector** | 0.43 | 3.75 | 5 |
| **7** | **Psychometric/Behavioral** | 0.35 | 2.00 | 1 |
| **8** | **Product/Tech** | 0.14 | 5.00 | 1 |

The verified results reveal a powerful and tightly clustered quartet of foundational predictors. The most significant predictor family is Firm Characteristics (WIS: 0.93), which is distinguished by having the highest feature count (12) while maintaining a strong average rank (2.75). This indicates that the fundamental, structural attributes of a venture—its age, location, and size—are the most consistently powerful and frequently cited predictors of success in the literature.

Following just behind are Investor _Structure (WIS: 0.92) and Digital and Social Traction (WIS: 0.90). Both demonstrate exceptional importance through very strong average ranks (2.60 and 2.56, respectively) and a high volume of evidence. The fourth member of this leading group is Funding History (WIS: 0.86), which also boasts a high feature count. Together, these top four families suggest that prediction models rely most heavily on a combination of a startup's structural facts, the quality of its investors, its online momentum, and its fundraising track record.

The classic "bet on the jockey" category, Team/Founder attributes (WIS: 0.75), is the first to show a notable drop-off in importance, though it remains a crucial and frequently cited factor. The least impactful categories in our analysis were high-level Market/Sector information and specific Product/Tech features, both of which were cited far less frequently in the top ranks.

## 3.3. Moderator Analysis: The Context-Dependency of Predictor Importance

While the overall ranking provides a general hierarchy, the true utility for investors and founders lies in understanding how predictor importance shifts across different contexts. To answer our second research question—"How does the importance of these predictors change depending on the context of the study?"—. We explored three key moderators: the definition of success, the startup's developmental stage, and the primary data source. This analysis reveals that there is no single, universal formula for success; rather, the most influential factors are highly context-dependent. The detailed feature importance ranks for each of the 13 studies, which form the basis for this moderator analysis, are provided in Appendix C.

### 3.3.1. The Moderating Effect of Success Outcome

Perhaps the most critical contextual factor is the definition of "success" itself. We compared studies predicting a definitive Exit Event (N=5) with those predicting a nearer-term Funding Milestone (N=3). Table 4 reveals a significant divergence in what matters most for these two distinct goals.

**Table 4**: Predictor Importance by Success Outcome (Ordered by WIS)

| Rank | Predicting `Exit Event` | WIS | Rank | Predicting `Funding Milestone` | WIS |
|---|---|---|---|---|---|
| **1** | Firm Characteristics | 0.86 | 1 | Funding History | 0.81 |
| **2** | Investor Structure | 0.59 | 2 | Market/Sector | 0.69 |
| **3** | Digital and Social Traction | 0.59 | 3 | Investor Structure | 0.55 |

For predicting a long-term Exit Event, fundamental, structural factors are paramount. Firm Characteristics (WIS: 0.86) emerges as the most important family, a finding strongly supported by a high count of 9 distinct features in the top ranks. This suggests that the structural integrity, age, and location of a company are critical for its ultimate viability. The next most important factors are the quality of the Investor Structure (WIS: 0.59) and the startup's ability to maintain Digital and Social Traction (WIS: 0.59).

In stark contrast, when predicting a nearer-term Funding Milestone, the focus shifts to the specifics of the deal itself. Funding History (WIS: 0.81), supported by a solid evidence base of 5 features, is the most important category. This highlights the importance of the current ask and prior funding events. The high score of Market/Sector (WIS: 0.69), though based on a single feature, underscores the importance of immediate market positioning in securing the next round of investment.

### 3.3.2. The Moderating Effect of Startup Stage

Next, we bifurcated our sample into Early Stage ventures (N=7) and General/Mixed-Stage ventures (N=6). Table 5 shows a clear shift in priorities as startups mature.

**Table 5**: Predictor Importance by Startup Stage (Ordered by WIS)

| Rank | `Early Stage` Models | WIS | Rank | `General/Mixed-Stage` Models | WIS |
|---|---|---|---|---|---|
| **1** | Funding History | 0.82 | 1 | Firm Characteristics | 0.86 |
| **2** | Team/Founder | 0.75 | 2 | Digital and Social Traction | 0.80 |
| **3** | Market/Sector | 0.69 | 3 | Investor Structure | 0.80 |

In Early Stage models, where performance data is limited, the analysis prioritizes the most tangible signals available. Funding History (WIS: 0.82), representing the specifics of the current or most recent deal, is the most important factor. This is followed by the Team/Founder (WIS: 0.75), which is supported by the largest volume of evidence in this group (10 features), confirming the classic "bet on the jockey" heuristic for new ventures.

For General/Mixed Stage models, the focus shifts decisively to more established, structural elements. Firm Characteristics (WIS: 0.86) becomes the clear leader, supported by the highest feature count (9). As companies grow, their fundamental structure becomes the most reliable predictor. This is followed by a tie between Digital and Social Traction (WIS: 0.80) and Investor Structure (WIS: 0.80), indicating that as startups mature, their ability to maintain online momentum and the quality of their investor syndicate are equally critical signals.

### 3.3.3. The Moderating Effect of Data Source

Finally, we analyzed the influence of the primary data source, comparing studies using Venture Database sources (N=6) with those using Bespoke Datasets (N=6).

**Table 6**: Predictor Importance by Primary Data Source (Ordered by WIS)

| Rank | **Venture Database** Models | WIS | Rank | **Bespoke Dataset** Models | WIS |
|---|---|---|---|---|---|
| **1** | Firm Characteristics | 0.86 | 1 | Funding History | 0.83 |
| **2** | Investor Structure | 0.80 | 2 | Team/Founder | 0.73 |
| **3** | Digital and Social Traction | 0.80 | 3 | Market/Sector | 0.69 |

This analysis reveals a powerful "lens" effect. Studies relying on large Venture Databases find the most importance in the data that is most structured and scalable within those platforms. Firm Characteristics (WIS: 0.86), with the highest feature count, is the top predictor, followed by a tie between the rich data on Investor Structure (WIS: 0.80) and Digital and Social Traction (WIS: 0.80).

Conversely, studies that create Bespoke Datasets find the most importance in the deep, nuanced data they are designed to capture. Funding History (WIS: 0.83) and Team/Founder attributes (WIS: 0.73) are the most important families, both supported by a high volume of features (9 and 8, respectively). This demonstrates that when researchers manually collect data, they prioritize information about the deal and the people behind it, and their models reflect this focus.

# 4. Discussion

The results of our meta-analysis provide a data-driven hierarchy of startup success predictors, but their interpretation requires a critical lens. In this section, we move beyond the quantitative findings to discuss their deeper implications. We first interpret the overall hierarchy, arguing that it reflects a systemic "convenience bias" in the literature. We then unpack the nuances revealed by our moderator analysis, highlighting the critical role of context. Finally, we offer a meta-scientific view on the state of the field and propose a path forward.

## 4.1. Interpreting the Main Findings: A New—and Biased—Hierarchy of Predictors

Our meta-analysis sought to answer a fundamental question: "What matters most for startup success?" The quantitative results in Section 3.2 highlight a dominant top-tier of predictor families: Firm Characteristics, Investor Structure, Digital and Social Traction, and Funding History. On the surface, this suggests a "four-pillar" framework for evaluation, a conclusion broadly supported by recent reviews like Sevilla-Bernardo et al. (2022), who identified similar factors. However, we argue that this hierarchy reflects not only an objective reality but also a systemic "convenience bias" in current research methodologies.

The prominence of these four families is understandable. They are composed of features that are highly structured, easily quantifiable, and readily available in the large-scale venture databases like Crunchbase that underpin much of the research. Data on a company's location (Firm Characteristics), total funding (Funding History), backer prestige (Investor Structure), and social media footprint (Digital and Social Traction) are straightforward to collect and analyze. This accessibility, however, can introduce significant bias. As Park et al. (2024) argue, a heavy reliance on post-inception data, such as funding events, can lead to a look-ahead bias that distorts the prediction of true Early Stage potential. Our findings therefore confirm that these structured data points are indeed powerful predictors, but we caution that their dominance may partly stem from their ease of access rather than their absolute, context-free importance.

The relatively lower rank of the Team/Founder family (WIS: 0.75) directly challenges the venture capital adage to "bet on the jockey." We posit that this ranking reflects not a lesser importance but rather a profound measurement difficulty. The features currently used in the literature are often simplistic proxies—number of founders, university rank, prior exits—that are easily extracted from platforms like Crunchbase or LinkedIn. These proxies fail to capture the deeper, more impactful founder attributes. Recent studies powerfully underscore this gap. Freiberg and Matz (2023) found that core personality traits like openness and agreeableness are highly predictive of funding success, while McCarthy et al. (2023) demonstrated that founder personalities significantly influence startup outcomes. This suggests that the next frontier lies in analyzing deeper, unstructured data—such as founder interviews, pitch deck sentiment, or team communication patterns—which can now be unlocked by LLM-based AI agents.

A second profound omission in the current literature, confirmed by our meta-analysis, is the near-total absence of macro-environmental features. None of the 13 studies in our sample incorporated macroeconomic indicators like GDP growth, interest rates, or regulatory changes.

This is a critical blind spot. As Díaz-Santamaría and Bulchand-Gidumal (2021) found, the economic context and location significantly influence startup success. By focusing almost exclusively on firm- and founder-level data, the field is building models that are context-agnostic and thus of limited real-world applicability for navigating different economic cycles.

Finally, our analysis reveals a widespread definitional imprecision in the field's central task. Many studies claim to predict "startup success" but use features only available to later-stage companies (e.g., extensive funding histories) to generate their predictions, creating a clear methodological flaw. As Argaw and Liu (2024) highlight, success definitions themselves vary widely by industry and context. For predictions to be truly valuable, especially at the early stages, models must be built exclusively on pre-funding data. The challenge, as Park et al. (2024) suggest through their work on bias-free machine learning, is not just to find what predicts success, but to find what predicts it *at the right time*. This represents the true frontier for next-generation AI in venture evaluation.

## 4.2. Unpacking the Nuances: The Critical Role of Context

While our aggregate analysis provides a general hierarchy of predictors, the most actionable insights emerge from our moderator analysis, which reveals how this hierarchy shifts depending on a study's specific context. Our findings powerfully demonstrate that there is no "one-size-fits-all" formula for prediction. The factors that matter most are contingent on the startup's goals, its developmental stage, and the very data used to evaluate it.

**The Journey vs. The Destination: Funding Milestones vs. Final Exits**
Our most striking finding is the dramatic difference between predicting a near-term funding event versus a long-term exit. When the goal is to secure a Funding Milestone, our analysis shows that predictors related to the immediate deal context—Funding History and Market/Sector—are paramount. This aligns with the practical reality of Early Stage pitching, where investors must make rapid decisions based on tangible, immediate signals. The findings of Böttcher et al. (2021), who studied Early Stage financing, support this view. They found that specific business models (a Market/Sector attribute) like "Two-Sided Market" and "Freemium" were significantly correlated with the amount of seed funding received, reinforcing the idea that immediate market positioning is critical for securing initial capital.

In stark contrast, when predicting a definitive Exit Event, the model's focus shifts to more fundamental, long-term indicators of viability. The quality of the Investor Structure and foundational Firm Characteristics become the most important predictors. This provides strong quantitative evidence for a two-stage logic in venture evaluation: investors assess a startup's immediate positioning to grant a *funding milestone*, but they rely on its structural integrity and the quality of its syndicate to predict a final, successful *exit*.

**From Seed to Oak: The Evolving Nature of Predictors**
Our analysis of the Startup Stage moderator further reinforces this temporal logic. For Early Stage ventures, where historical performance data is non-existent, models prioritize proxies for potential. Our WIS metric shows that Funding History and Team/Founder attributes are most critical. This confirms the conventional wisdom that Early Stage investment is fundamentally a bet on the deal and the people. The seminal survey by Gompers et al. (2016) provides direct support for this finding, reporting that a staggering 95% of VCs rate the management team as an important factor, with 47% citing it as the single *most* important factor in their investment decisions.

Conversely, as startups mature into the General/Mixed-Stage category, the predictive focus shifts. Firm Characteristics becomes the top predictor, followed closely by Digital and Social Traction and Investor Structure. This suggests that for a growing company, its established structure (size, age), its ability to maintain market momentum, and the continued support of a strong investor syndicate become the most telling signs of future success, moving beyond the initial assessment of the founding team.

**The Lens Determines the View: A Meta-Scientific Insight on Data Sources**
Finally, our analysis of the Data Source moderator offers a critical insight into the research process itself. The "most important" features are often an artifact of the data source used. Studies anchored in large-scale Venture Databases find the most predictive power in the highly structured data these platforms provide: Firm Characteristics, Investor Structure, and Digital and Social Traction. In contrast, studies that build Bespoke Datasets—like the Gompers et al. (2016) survey or the targeted web scraping in Antretter et al. (2019)—find the most importance in the deep, nuanced data they were designed to collect: Funding History (the specifics of a deal) and Team/Founder attributes. This does not mean one group is "right" and the other "wrong." Rather, it proves that the choice of data acts as a lens that brings certain types of predictors into focus. This finding is a powerful argument for the necessity of data fusion, combining the scale of venture databases with the depth of bespoke data to achieve a more holistic and accurate view of the drivers of startup success.

## 4.3. A Meta-Scientific View: Limitations, Biases, and a Call to Action

While our meta-analysis provides a robust synthesis of the existing literature, its findings must be understood within the context of its own limitations, many of which reflect systemic biases and gaps in the field of AI-driven startup prediction. By examining these challenges, we can propose a clear agenda for maturing the discipline from a collection of disparate studies into a more cohesive body of scientific knowledge.

First, our own analysis is constrained by the available evidence. The final sample of N=13 studies, while representing the entirety of the literature that reports rankable feature importance, is still relatively small. This limits the statistical power of our moderator analyses and means that some of our contextual findings, while directionally insightful, should be interpreted as strong hypotheses for future verification rather than definitive laws. Furthermore, our categorization of 58 unique features into eight "Predictor Families" necessarily involves a degree of scholarly interpretation. While we have documented this process transparently (see Appendices A and B), it represents one possible valid taxonomy.

Second, our meta-analysis exposes a significant publication and data availability bias in the literature. The near-total absence of Psychometric/Behavioral features in our results is not because they are unimportant—indeed, recent work by Freiberg & Matz (2023) and McCarthy et al. (2023) proves they are critically important—but because they are exceptionally difficult to collect at scale. This leads to a field that over-emphasizes what is easy to measure (e.g., funding amounts from Crunchbase) rather than what may be most fundamental. This creates a risk that the literature is producing highly-tuned models based on convenient, but potentially secondary, signals, while ignoring the deeper, causal drivers of success.

This leads to our final point and a call to action for the research community. For this field to advance scientifically, a greater commitment to methodological transparency and standardized reporting is required. Our review process revealed numerous high-quality studies that could not be included in this meta-analysis simply because they did not report their feature importances. This is a missed opportunity for collective knowledge-building. We propose that journals and premier conferences in this domain should strongly encourage, if not mandate, that empirical papers adhere to a simple set of reporting standards:

1. **Report Feature Importance:** Alongside model performance metrics like AUC or accuracy, studies should always include a full, ranked list or chart of the feature importances that drive their predictions.
2. **Provide Unambiguous Definitions:** Studies must clearly define their success metric (e.g., "success is defined as an IPO or acquisition >$50M") and the exact data sources used, including the time period of data collection.
3. **Encourage Data and Code Sharing:** Whenever possible, sharing the underlying feature set and code allows for replication and extension, which is the cornerstone of scientific progress.

Adopting these standards will not only improve the quality of individual studies but will enable future meta-analyses like ours to be conducted on a much larger scale, providing more powerful, robust, and granular insights. By doing so, we can collectively accelerate the maturation of AI in startup evaluation from an art subsidized by convenience to a true, replicable science.

# 5. Conclusion

The proliferation of data and machine learning has opened new frontiers for understanding the complex dynamics of startup success. However, it has also produced a fragmented landscape of findings, making it difficult to discern which predictive signals are truly robust. This study undertook a comprehensive effort to bring clarity to this field. Our research began with a systematic literature review that identified and analyzed 57 relevant empirical studies. From this large corpus, we then isolated the specific subset of papers that provided the necessary data for a quantitative synthesis of predictor importance. While this rigorous filtering resulted in a final sample of 13 studies, they represent the entirety of the literature that currently allows for such an analysis, providing a unique and valuable lens into the field's collective findings.

### 5.1. Summary of Contributions

Our research makes three primary contributions. First, by synthesizing the feature importance rankings from the qualifying literature, we established the first evidence-based hierarchy of predictor families. Our analysis reveals that the most consistently powerful predictors are a quartet of foundational attributes: the venture's fundamental Firm Characteristics, the quality of its Investor Structure, its real-time Digital and Social Traction, and its proven Funding History.

Second, and perhaps more importantly, we demonstrated that this hierarchy is not static. Our moderator analysis provides strong quantitative evidence that "what matters most" is highly context-dependent. The signals that predict a near-term Funding Milestone are fundamentally different from those that predict a long-term Exit Event, challenging the notion of a universal prediction model and arguing for a more nuanced, context-aware approach to startup evaluation.

Third, our study offers a meta-scientific critique of the field. We identified a systemic "convenience bias," where the dominance of certain predictors is tied to their ease of measurement in large-scale databases, and a widespread lack of macroeconomic context in current models. In doing so, we have not only synthesized the existing research but have also highlighted the critical gaps that must be addressed for the field to mature.

### 5.2. Limitations and Future Research

This study, like all research, has its limitations, which in turn illuminate promising avenues for future inquiry. The fact that only 13 of 57 relevant studies provided rankable feature importance is a limitation of the field itself, but it restricts the statistical power of our moderator analyses. As more studies adopt transparent reporting practices, future meta-analyses will be able to draw more robust conclusions.

The gaps identified by our analysis point toward a clear research agenda. There is a critical need for studies that move beyond convenient, structured data, particularly by incorporating Psychometric/Behavioral traits of founders and macroeconomic indicators. Furthermore, the "lens effect" of data sources demonstrated by our moderator analysis is a powerful argument for the superiority of data fusion approaches. The rise of LLM-powered AI agents capable of parsing unstructured data will be the key enabler of this next wave of research.

In conclusion, this meta-analysis provides a structured synthesis of what we currently know about AI-driven startup prediction while simultaneously charting a course for what we still need to learn. By moving toward more context-aware models and committing to greater methodological transparency, we can advance this field from a collection of individual insights into a cohesive body of scientific knowledge.

# References

1. Antretter, T., Blohm, I., Grichnik, D., & Wincent, J. (2019). Predicting new venture survival: A Twitter-based machine learning approach to measuring online legitimacy. *Journal of Business Venturing Insights*, *11*, e00109. https://doi.org/10.1016/j.jbvi.2018.e00109

2. Argaw, Y. M., & Liu, Y. (2024). The pathway to startup success: A comprehensive systematic review of critical factors and the future research agenda in developed and emerging markets. *Systems*, *12*(12), 541. https://doi.org/10.3390/systems12120541

3. Böttcher, T. P., Bootz, V., Zubko, T., Weking, J., Böhm, M., & Krcmar, H. (2021, March). Enter the shark tank: the impact of business models on early stage financing. In *International Conference on Wirtschaftsinformatik* (pp. 275-289). Cham: Springer International Publishing.

4. Cholil, S. R., Gernowo, R., Widodo, C. E., Wibowo, A., Warsito, B., & Hirzan, A. M. (2024). Predicting startup success using tree-based machine learning algorithms. *Revista de Informática Teórica e Aplicada*, *31*(1), 50–59. https://doi.org/10.22456/2175-2745.133375

5. Deodhar, O., Bhatkar, S., Dixit, P., Kulkarni, P., Oak, A., & Londhe, S. (2024). Shark tank deal prediction using machine learning techniques. In *2024 IEEE 9th International Conference for Convergence in Technology (I2CT)* (pp. 1–6). IEEE. https://doi.org/10.1109/I2CT61223.2024.10543650

6. Díaz-Santamaría, C., & Bulchand-Gidumal, J. (2021). Econometric estimation of the factors that influence startup success. *Sustainability*, *13*(4), 2242. https://doi.org/10.3390/su13042242

7. Dziubanovska, N., Maslii, V., Shekhanin, O., & Protsyk, S. (2024). Machine learning for assessing startup investment attractiveness. In *The First International Workshop of Young Scientists on Artificial Intelligence for Sustainable Development, May 10-11, 2024, Ternopil, Ukraine* (Vol. 3716). CEUR-WS.org.

8. Eljil, K. S., & Naït-Abdesselam, F. (2024). Predicting initial coin offering success using machine learning and human capital analysis. In *2024 IEEE/ACM International Conference on Big Data Computing, Applications and Technologies (BDCAT)* (pp. 348-353). IEEE. https://doi.org/10.1109/BDCAT63179.2024.00060

9. Ferry, T., Goyal, M., Sidhu, I., & Fred-Ojala, A. (2018). Breaking the Zuckerberg myth: Successful entrepreneurs have 10 years of prior employment. In *2018 IEEE Technology & Engineering Management Conference (TEMSCON)* (pp. 1-6). IEEE. https://doi.org/10.1109/ICE.2018.8436327

10. Freiberg, B., & Matz, S. C. (2023). Founder personality and entrepreneurial outcomes: A large-scale field study of technology startups. *Proceedings of the National Academy of Sciences*, *120*(18), e2215829120. https://doi.org/10.1073/pnas.2215829120

11. Gangwani, D., Zhu, X., & Furht, B. (2023). Exploring investor-business-market interplay for business success prediction. *Journal of Big Data*, *10*(1), 48. https://doi.org/10.1186/s40537-023-00723-6

12. Gautam, L., & Wattanapongsakorn, N. (2024). Machine learning models to investigate startup success in venture capital using Crunchbase dataset. In *2024 21st International Joint Conference on Computer Science and Software Engineering (JCSSE)* (pp. 475-481). IEEE. https://doi.org/10.1109/JCSSE61278.2024.10613650

13. Gompers, P. A., Gornall, W., Kaplan, S. N., & Strebulaev, I. A. (2016). *How Do Venture Capitalists Make Decisions?* (NBER Working Paper No. 22587). National Bureau of Economic Research. https://doi.org/10.3386/w22587

14. Hu, J., Hu, L., Hu, M., & He, Q. (2022). Machine learning-based investigation on the impact of Chinese venture capital institutions' performance: Evaluation factors of venture enterprises to venture capital institutions. *Systems*, *10*(4), 92. https://doi.org/10.3390/systems10040092

15. Kim, J., Kim, H., & Geum, Y. (2023). How to succeed in the market? Predicting startup success using a machine learning approach. *Technological Forecasting and Social Change*, *193*, 122614. https://doi.org/10.1016/j.techfore.2023.122614

16. McCarthy, P. X., Gong, X., Braesemann, F., Kern, M. L., & Rizoiu, M.-A. (2023). The impact of founder personalities on startup success. *Scientific Reports*, *13*(1), 17200. https://doi.org/10.1038/s41598-023-41980-y

17. Moterased, M., Sajadi, S. M., Davari, A., & Zali, M. R. (2021). Toward prediction of entrepreneurial exit in Iran; a study based on GEM 2008-2019 data and approach of machine learning algorithms. *Big Data and Computing Visions*, *1*(3), 111-127. https://doi.org/10.22105/BDCV.2021.142089

18. Page, M. J., McKenzie, J. E., Bossuyt, P. M., Boutron, I., Hoffmann, T. C., Mulrow, C. D., Shamseer, L., Tetzlaff, J. M., Akl, E. A., Brennan, S. E., Chou, R., Glanville, J., Grimshaw, J. M., Hróbjartsson, A., Lalu, M. M., Li, T., Loder, E. W., Mayo-Wilson, E., McDonald, S., … Moher, D. (2021). The PRISMA 2020 statement: An updated guideline for reporting systematic reviews. *BMJ*, *372*, n71. https://doi.org/10.1136/bmj.n71

19. Park, J., Choi, S., & Feng, Y. (2024). Predicting startup success using two bias-free machine learning: Resolving data imbalance using generative adversarial networks. *Journal of Big Data*, *11*(1), 118. https://doi.org/10.1186/s40537-024-00993-8

20. Razaghzadeh Bidgoli, M., Vanani, I. R., & Goodarzi, M. (2024). Predicting the success of startups using a machine learning approach. *Journal of Innovation and Entrepreneurship*, *13*(1), 80. https://doi.org/10.1186/s13731-024-00436-x

21. Sevilla-Bernardo, J., Sanchez-Robles, B., & Herrador-Alcaide, T. C. (2022). Success factors of startups in research literature within the entrepreneurial ecosystem. *Administrative Sciences*, *12*(3), 102. https://doi.org/10.3390/admsci12030102

22. Żbikowski, K., & Antosiuk, P. (2021). A machine learning, bias-free approach for predicting business success using Crunchbase data. *Information Processing & Management*, *58*(4), 102555. https://doi.org/10.1016/j.ipm.2021.102555

## Appendix A: Feature Importance Mapping by Source Study

This appendix details the specific features extracted from each of the 13 studies included in the meta-analysis. For each study, this table maps its five most important features to their corresponding standardized predictor family.

**Legend:**

- ★ **(Top 3):** Indicates a feature from this family was ranked 1st, 2nd, or 3rd.
- ✓ **(Top 5):** Indicates a feature from this family was ranked 4th or 5th.

**Table A: Mapping of Top 5 Features to Predictor Families**

| ***Source Paper*** | ***Team/ Founder*** | ***Digital and Social Traction*** | ***Firm Characteristics*** | ***Funding History*** | ***Investor Structure*** | ***Market/ Sector*** | ***Product/ Tech*** | ***Psychometric*** |
|---|---|---|---|---|---|---|---|---|
| ***Antretter (2019)*** | | ★, ★, ★, ✓, ✓ | | | | | | |
| ***Cholil (2024)*** | | | ★ | ★, ★, ✓, ✓ | | | | |
| ***Deodhar (2024)*** | | | | ★, ✓ | | ★ | | |
| ***Dziubanovska (2024)*** | | | | ★, ✓ | ★ | | | |
| ***Eljil (2024)*** | ★, ✓, ✓ | | | ★ | ★ | | | |
| ***Ferry (2018)*** | ★, ★, ★, ✓, ✓ | | | | | | | |
| ***Gangwani (2023)*** | | | ★ | | ★, ★, ✓ | ✓ | | |
| ***Gautam (2024)*** | | ★ | ★, ★,✓ | | | | ✓ | |
| ***Hu (2022)*** | | | | ★ | ★, ★, ★, ✓, ✓ | ★ | | |
| ***Kim (2023)*** | | ★ | ★ | ★ | | ✓, ✓ | | |
| ***Moterased (2021)*** | | | | | | | | ★ |
| ***Razaghzadeh (2024)*** | | ★,✓ | ★, ★,✓ | | | | | |
| ***Żbikowski (2021)*** | ★, ✓ | | ★, ★,✓ | | | | | |
| ***TOTALS:*** | | | | | | | | |
| ***Count in Top 3*** | **5** | **6** | **9** | **7** | **7** | **2** | **0** | **1** |
| ***Count in Top 5*** | **10** | **9** | **12** | **11** | **10** | **5** | **1** | **1** |

---

## Appendix B: Full Feature Rank Ledger for Average Rank Calculation

This table provides the complete dataset used to calculate the "Average Rank" for each predictor family.

**Table B**: Comprehensive Ledger of All Ranked Features

| *Source Paper* | *Rank* | *Feature Name* | *Coded Predictor Family* |
|---|---|---|---|
| *Antretter (2019)* | 1 | Length of tweets | Digital and Social Traction |
| | 2 | No. of likes given | Digital and Social Traction |
| | 3 | No. of likes received | Digital and Social Traction |
| | 4 | Followers per following | Digital and Social Traction |
| | 5 | No. of followings | Digital and Social Traction |
| *Cholil (2024)* | 1 | age_startup_year | Firm Characteristics |
| | 2 | last_milestone_year | Funding History |
| | 3 | first_funding_year | Funding History |
| | 4 | age_last_milestone | Funding History |
| | 5 | age_first_milestone | Funding History |
| *Deodhar (2024)* | 1 | Category | Market/Sector |
| | 2 | pitcher_ask_amount | Funding History |
| | 3 | ask_equity | Funding History |
| *Eljil (2024)* | 1 | Goal | Funding History |
| | 2 | ICO_amount_team | Team/Founder |
| | 3 | ICO_amount_advisor | Investor Structure |
| | 4 | Team_avg_exp | Team/Founder |
| | 5 | Verified (Team) | Team/Founder |
| *Ferry (2018)* | 1 | Years of Employment | Team/Founder |
| | 2 | Standardized Major | Team/Founder |
| | 3 | Standardized Graduate Institution | Team/Founder |
| | 4 | Degree Type | Team/Founder |
| | 5 | Standardized University | Team/Founder |
| *Gangwani (2023)* | 1 | Percentage of Success Rate - I | Investor Structure |
| | 2 | Percentage of Failure Rate - I | Investor Structure |
| | 3 | Number of Successful Companies - B | Firm Characteristics |
| | 4 | Sum of Successful Raised Amount - I | Investor Structure |
| | 5 | Top Current Sector - M | Market/Sector |
| *Gautam (2024)* | 1 | is_Org_Home_Page_Active | Digital and Social Traction |
| | 2 | rank | Firm Characteristics |
| | 3 | age_of_org_in_days | Firm Characteristics |
| | 4 | age_of_org_in_years | Firm Characteristics |
| | 5 | short_description_bag_or_word | Product/Tech |
| *Hu (2022)* | 1 | Investment Size | Investor Structure |
| | 2 | LPN (Number of limited partners) | Investor Structure |
| | 3 | Exit Type | Investor Structure |
| | 4 | NFM (Number of funds) | Investor Structure |
| | 5 | EN (Exit number) | Investor Structure |
| *Kim (2023)* | 1 | Employees_low | Firm Characteristics |
| | 2 | Exposure | Digital and Social Traction |
| | 3 | Funding Monetary | Funding History |
| | 4 | Industry convergence level | Market/Sector |
| | 5 | Industry association level | Market/Sector |
| *Moterased (2021)* | 1 | exreason | Other |
| | 2 | bstart | Psychometric/Behavioral |
| *Razaghzadeh (2024)* | 1 | LinkedIn Followers | Digital and Social Traction |
| | 2 | Timelapse_until_fifth_year | Firm Characteristics |
| | 3 | Employs (on LinkedIn) | Firm Characteristics |
| | 4 | Twitter Followers | Digital and Social Traction |
| | 5 | startup_age | Firm Characteristics |
| *Żbikowski (2021)* | 1 | country_code | Firm Characteristics |
| | 2 | has_multiple_degrees | Team/Founder |
| | 3 | region_size | Firm Characteristics |
| | 4 | gender_male | Team/Founder |
| | 5 | city_size | Firm Characteristics |
| *Dziubanovska (2024)* | 1 | Backers | Investor Structure |
| | 2 | Goal | Funding History |
| | 3 | Pledged | Funding History |

## Appendix C: Feature Importance Profiles by Source Study

This appendix provides the detailed data used to calculate the average ranks for all moderator analyses. For each of the 13 studies in our meta-analytic sample, this table shows the rank of each of its Top 5 most important features, organized by its standardized predictor family. An empty cell indicates that no feature from that family appeared in the Top 5 for that study. This structure allows for easy verification of all group-level calculations.

**Table C**: Rank of Top 5 Features by Family for Each Source Study

| *Source Paper (First Author)* | *Team/ Founder* | *Digital and Social Traction* | *Firm Characteristics* | *Funding History* | *Investor Structure* | *Market/ Sector* | *Product/ Tech* | *Psychometric* |
|---|---|---|---|---|---|---|---|---|
| *Antretter (2019)* | | 1, 2, 3, 4, 5 | | | | | | |
| *Cholil (2024)* | | | 1 | 2, 3, 4, 5 | | | | |
| *Deodhar (2024)* | | | | 2, 3 | | 1 | | |
| *Dziubanovska (2024)* | | | | 2, 3 | 1 | | | |
| *Eljil (2024)* | 2, 4, 5 | | | 1 | 3 | | | |
| *Ferry (2018)* | 1, 2, 3, 4, 5 | | | | | | | |
| *Gangwani (2023)* | | | 3 | | 1, 2, 4 | 5 | | |
| *Gautam (2024)* | | 1 | 2, 3, 4 | | | | 5 | |
| *Hu (2022)* | | | | | 1, 2, 3, 4, 5 | | | |
| *Kim (2023)* | | 2 | 1 | 3 | | 4, 5 | | |
| *Moterased (2021)* | | | | | | | | 2 |
| *Razaghzadeh (2024)* | | 1, 4 | 2, 3, 5 | | | | | |
| *Żbikowski (2021)* | 2, 4 | | 1, 3, 5 | | | | | |